\documentclass[aps,prl,twocolumn,superscriptaddress,showpacs,floatfix]{revtex4-2}

\usepackage{graphicx}
\usepackage{amsmath}
\usepackage{amssymb}
\usepackage{hyperref}

\newcommand{\Tr}{\mathrm{Tr}}
\begin{document}

\title{Mean-Force Hamiltonians from Influence Functionals}
\author{Gerard McCaul}
\date{\today}

\begin{abstract}
The Hamiltonian of mean force (HMF) provides the standard starting point for strong-coupling thermodynamics, yet explicit operator forms are known only in restricted settings. We present a ``quenched density" framework that uses the Hubbard-Stratonovich transformation to rewrite the reduced equilibrium state as an average over local propagators in imaginary time. This approach rigorously separates the statistical definition of the environment from the algebraic structure of the system response. We apply this framework to the minimal case of a harmonic environment with a coupling commuting with the system Hamiltonian. In this scenario the correction to the HMF has an exact, closed-form expression. We validate this result against finite-bath trace-out calculations and stochastic imaginary-time sampling in a five-level projector-coupled model. This confirms the specific strong-coupling projector dominance predicted by energetic arguments 
\end{abstract}

\maketitle

\section{\label{sec:intro}Introduction}

What governs equilibrium? The canonical answer is the thermal Gibbs state, where the environment is characterised by a single parameter: the temperature. Yet, this canonical answer is puzzling. It implies that neither the specific character of the environment nor the form of its coupling to the system alters the system's equilibrium. This of course is a consequence of the assumptions used to derive the answer - the weak-coupling or Markovian regimes open-systems theory has historically relied upon \cite{breuerTheoryOpenQuantum2002}. At strong coupling however, the reduced state inherits explicit temperature dependence and interaction-induced operator content that definitionally cannot be captured by these assumptions \cite{hanggiFiniteQuantumDissipation2008,ingoldSpecificHeatAnomalies2009}. These issues have motivated a broad investigation into strong-coupling thermodynamics, including exact fluctuation relations~\cite{campisiFluctuationTheoremArbitrary2009,jarzynskiNonequilibriumWorkTheorem2004,gooldNonequilibriumThermodynamics2019}, consistent heat and work definitions~\cite{espositoNatureHeatStrongly2015,rivasStrongCouplingThermodynamics2020,lacerdaQuantumThermodynamicsFast2023}, and operational measurability~\cite{strasbergMeasurabilityNonequilibriumThermodynamics2020,millerEntropyProduction2017,binderOperationalThermodynamics2019}. This representability question sits alongside operational ambiguities in thermodynamic bookkeeping beyond weak coupling, where heat/work/state assignments can depend on the chosen framework~\cite{BinderVinjanampathyModiGoold2015}. 

While formally one can write the reduced state in canonical exponential form, the effective Hamiltonian governing this state is generally not the bare Hamiltonian of the system. Instead, it must account for the non-trivial influence of the environment in a manner that goes beyond the specification of temperature. If an effective operator capturing this influence can be constructed, the standard description of equilibrium holds generically across all coupling strengths. \emph{The Hamiltonian of mean force} (HMF) is the operator that reproduces this reduced equilibrium object as a Gibbs-like state, and provides the standard starting point for strong-coupling thermodynamics\cite{campisiFluctuationTheoremArbitrary2009,jarzynskiNonequilibriumWorkTheorem2004,jarzynskiStochasticMacroscopicThermodynamics2017,talknerColloquiumStatisticalMechanics2020,seifertFirstSecondLaw2016}. Historically, much of open-system theory emphasised weak-coupling and Markovian regimes, where reduced equilibrium states are well approximated by Gibbs states of renormalized system Hamiltonians. At finite coupling, equilibrium consistency is anchored in the correlated global Gibbs state, and the reduced operator inherits coupling-dependent structure that need not be captured by a simple renormalization. The HMF can become temperature dependent and can encode interaction-induced terms that are absent in $H_Q$, including effective many-body or nonlocal operator content\cite{talknerColloquiumStatisticalMechanics2020,seifertFirstSecondLaw2016,espositoNatureHeatStrongly2015,hanggiFiniteQuantumDissipation2008,ingoldSpecificHeatAnomalies2009,correaPotentialRenormalisationLamb2025}. This is the conceptual tension: the reduced equilibrium state is well defined, but its generator is not generally simple. In the strong-coupling regime one instead characterises equilibrium through the mean-force Gibbs state, whose structure and limiting forms (including ultrastrong-coupling projector dominance) have been analysed from a `state-first' perspective~\cite{cresserWeakUltrastrongCoupling2021a,trushechkinOpenQuantumSystem2022}. The present work is complementary: we start from the Euclidean influence functional and give an explicit local stochastic representation that yields a closed operator generator in a certified commuting--Gaussian sector. Related approaches connecting the HMF and influence functional have recently been explored in Refs.~\cite{hoggTutorialStochasticSimulation2024,hoggEnhancedEntanglement2024}. 

The ultimate target of thermodynamic descriptions is therefore unambiguous, and in fact a formal definition of HMF is equally unambiguous - it is simply the logarithm of the total state after tracing out its environment. It is the \emph{representation} of this operator that remains a challenge. Is the HMF a local operator? Does it admit a compact description in terms of few-body interactions? To address this, a veritable zoo of approximations - polaron transforms \cite{jangNonequilibriumTheoryElectronic2008}, weak-coupling expansions~\cite{cresserWeakUltrastrongCoupling2021a}, reaction-coordinate mappings~ \cite{ShubrookIlesSmithNazir2025, duGeneralizedHamiltonianMeanforce2025a,nazirReactionCoordinateMapping2018,AntoSztrikacsNazirSegal2023,IlesSmithDijkstraLambertNazir2016} - have proliferated strong-coupling thermodynamics. These approximations often disagree in intermediate regimes however, and while powerful numerical techniques have been derived from path-integrals~\cite{moixEquilibriumreducedDensityMatrix2012,chenRigorousStochasticMatrix2014,makriExploitingClassicalDecoherence2014,tanimuraReducedHierarchicalEquations2014,songCalculationCorrelatedInitial2015,WaltersWang2024}, they typically do not yield compact operator forms for the HMF.

Ultimately, the source of these difficulties lie in the trace operation, which mixes the bath and system degrees of freedom in a way that is difficult to invert. For this reason, a general framework for representing the HMF has proved elusive. The purpose of this paper is to lay the groundwork for just such a framework. To do so, we construct a representation of the equilibrium density matrix that disentangles the statistical properties of the environment from the algebraic response of the system. We term this object the \emph{quenched density}, and to obtain it we employ the \emph{influence functional} approach \cite{feynmanTheoryGeneralQuantum1963a,caldeiraQuantumTunnellingDissipative1983a,grabertQuantumBrownianMotion1988}. This formalism provides a natural language for the problem. For a recent unifying overview that links hierarchical, chain-mapping, pseudomode, stochastic, and extended-state-space approaches to non-Markovian open-system dynamics, see Ref.~\cite{XuVadimovStockburgerAnkerhold2026}. By employing the influence functional in concert with the Hubbard-Stratonovich decoupling~\cite{hubbardCalculationPartitionFunctions1959a,stratonovich1957QDistro,stockburgerExactNumberRepresentation2002}, we are able to rewrite the reduced equilibrium density not as a partial trace, but as a classical average over an ensemble of ``quenched'' imaginary-time propagators.
\begin{equation}
    \bar{\rho}_S(\beta) = \mathbb{E}_{\xi}\left[ \hat{\mathcal{T}} e^{-\int_0^\beta d\tau H_{\mathrm{eff}}[\xi(\tau)]} \right].
\end{equation}
In this picture, the bath is replaced by a stochastic field $\xi(\tau)$~\cite{stockburgerSimulatingSpinbosonDynamics2004,mccaulPartitionfreeApproachOpen2017c,wiedmannNonMarkovianDynamicsQuantum2020,laneExactlyThermalizedQuantum2020,matosEfficientChoiceColored2020,matosStochasticEntropyProduction2022}, and the system evolves under a time-dependent effective Hamiltonian for each realisation of the noise.

This formulation is obtained in two steps. The path integral sets the system in a frame where it can be expressed in terms of classical trajectories for the bath degrees of freedom, which can be integrated out. The resulting action is then a non-local kernel in imaginary time. The Hubbard-Stratonovich transformation remedies this by representing the bath information locally. This then allows the path integral to be reversed and the reduced density to be formulated at the operator level. The framework thus provides a rigorous bridge between the nonlocal path-integral description (where the bath is exact but the system operator structure is opaque) and the local operator description required for thermodynamics.

We apply this framework to establishing a rigorous benchmark: the linearly coupled Gaussian bath in the commuting (pure-dephasing) sector~\cite{clerkIntroductionQuantumNoise2010,breuerTheoryOpenQuantum2002}. In this limit, relevant for quantum nondemolition measurements and qubit dephasing~\cite{makhlinQuantumStateEngineering2001,palmaQuantumComputersDissipation1996}, the technical obstruction of imaginary time ordering vanishes, and the quenched density's average can be performed analytically. We use this recover the known result that the bath induces a static potential renormalisation $-\lambda f^2$ (the reorganisation energy)~\cite{campisiTalknerHanggi2009Solvable}. While this specific result is standard, its derivation within the quenched density framework serves two critical purposes. First, it demonstrates how the framework naturally recovers thermodynamic identities from path-integral primitives. Second, and most importantly, it provides a foundation for the generalisation of the framework to both non-commuting couplings and anharmonic environments.

The paper is structured as follows. In Sec.~\ref{sec:model} we define the total Hamiltonian and reduced equilibrium objects. Sec.~\ref{sec:quenched} develops the core of the framework: the Quenched Density representation of the imaginary-time propagator. Sec.~\ref{sec:influence_to_hmf} connects this to the exact commuting Gaussian benchmark. Sec.~\ref{sec:numerical} presents the numerical validation, confirming that the quenched density sampling, analytic prediction, and full finite-bath trace-out agree to machine precision. Sec.~\ref{sec:discussion} closes the paper with a summary of results, and highlights the essential contribution of this framework as a foundation for the future extension of the theory to non-Gaussian and non-commuting environments.
\section{\label{sec:model}The Total Hamiltonian and the Hamiltonian of Mean Force}
The Hamiltonian of mean force is defined by tracing the global equilibrium
operator, and therefore presupposes a concrete model for the composite
Hamiltonian. We consider a system with Hamiltonian $H_Q$ (system coordinates
$q$), coupled to a bath with Hamiltonian $H_X$ (bath coordinates $x$) through an
interaction $H_I$, so that
\begin{equation}
    H_{\mathrm{tot}}=H_Q+H_X+H_I.
    \label{eq:Htot_model}
\end{equation}
To obtain an explicit influence-functional representation we specialise to a
Gaussian environment with linear coupling, i.e.\ a Caldeira-Leggett bath \cite{caldeiraQuantumTunnellingDissipative1983a}. This
choice is appropriate when the bath is large and near equilibrium, so that its
fluctuations are well captured by a quadratic Hamiltonian and the leading
(system-local) term in the interaction dominates.

The bath is modeled as a collection of harmonic oscillators,
\begin{equation}
    H_X=\sum_k\left(\frac{p_k^2}{2m_k}+\frac{1}{2}m_k\omega_k^2 x_k^2\right),
    \label{eq:HX_model}
\end{equation}
and we take a linear coupling to a system operator $f\equiv f(q)$,
\begin{equation}
    H_I=\sum_k c_k f(q)\, x_k,
    \label{eq:HI_model}
\end{equation}
so that
\begin{equation}
    H_{\mathrm{tot}} = H_Q + \sum_k \left( \frac{p_k^2}{2m_k} + \frac{1}{2}m_k\omega_k^2 x_k^2 + c_k f(q)\, x_k \right).
    \label{eq:Htot_full}
\end{equation}
As is standard in Caldeira-Leggett constructions, one may include a counter-term
proportional to $f(q)^2$ to enforce translational invariance of the bath
coordinates and to compensate the static potential renormalisation induced by
the coupling \cite{caldeiraQuantumTunnellingDissipative1983a}. In what follows we absorb any such counterterm into the
definition of $H_Q$.

Although the microscopic bath is parametrised by $\{m_k,\omega_k,c_k\}$, its
influence on the system enters only through the spectral density~\cite{weissQuantumDissipativeSystems2012, breuerTheoryOpenQuantum2002},
\begin{equation}
    J(\omega) = \frac{\pi}{2} \sum_k \frac{c_k^2}{m_k \omega_k} \delta(\omega-\omega_k),
    \label{eq:spectral_density}
\end{equation}
which compactly encodes the bath two-point structure and fixes the imaginary-time influence kernel used below~\cite{feynmanTheoryGeneralQuantum1963a}.

Having specified the composite dynamics, we define the Hamiltonian of mean force by tracing the global Hamiltonian of mean force $H_{\mathrm{MF}}(\beta)$ as the operator whose Gibbs
form reproduces the reduced equilibrium object up to a chosen normalisation. If
the total system is at inverse temperature $\beta$, the unnormalised global
equilibrium operator is $e^{-\beta H_{\mathrm{tot}}}$, then the reduced
(unnormalised) equilibrium operator is
\begin{equation}
    \bar{\rho}_Q(\beta)\equiv \Tr_X\, e^{-\beta H_{\mathrm{tot}}},
    \label{eq:reduced_equilibrium_operator}
\end{equation}
where $\Tr_X$ traces out the bath degrees of freedom. The Hamiltonian of mean
force is defined by
\begin{equation}
    e^{-\beta H_{\mathrm{MF}}(\beta)}
    \equiv
    \frac{\Tr_X\, e^{-\beta H_{\mathrm{tot}}}}{Z_X(\beta)},
    \qquad
    Z_X(\beta)\equiv \Tr_X\, e^{-\beta H_X},
    \label{eq:HMF_def}
\end{equation}
or equivalently
\begin{equation}
H_{\mathrm{MF}}(\beta)=-(1/\beta)\log\!\big[\Tr_X e^{-\beta H_{\mathrm{tot}}}\big]
\end{equation}
up to an additive scalar fixed by the choice of $Z_X$.

For a harmonic bath linearly coupled through a collective coordinate, the bath
can be eliminated exactly, yielding the Feynman-Vernon influence functional
\cite{feynmanTheoryGeneralQuantum1963a, grabertQuantumBrownianMotion1988, ingoldSpecificHeatAnomalies2009, hanggiFiniteQuantumDissipation2008}.
In equilibrium this produces a Euclidean (imaginary-time) influence functional
which is bilocal in $\tau$ and whose kernel is fixed by the bath spectral
density $J(\omega)$~\cite{moixEquilibriumreducedDensityMatrix2012, mccaulPartitionfreeApproachOpen2017c, mccaulDrivingSpinbosonModels2018a, mccaulHowWinFriends2021b}. While this gives an exact representation of the reduced equilibrium operator, it does not generically yield a compact local expression for $H_{\mathrm{MF}}$. The reduced equilibrium object is a path integral over the system degrees of freedom, rather than a
simple exponential $e^{-\beta H_{\mathrm{eff}}}$ of a time-independent operator.
Extracting $H_{\mathrm{MF}}$ - the logarithm of this ordered object - is
therefore an inverse representability problem~\cite{ColemanYukalov2000RDM,Mazziotti2007RDMMechanics,LiuChristandlVerstraete2007NRepQMA,Klyachko2006QuantumMarginal}, not merely a question of coupling
strength or temperature~\cite{talknerColloquiumStatisticalMechanics2020, hovhannisyanHamiltonianMeanForce2020}.

Theere are two logically distinct aspects to this problem. First, the bath produces an \emph{imaginary-time nonlocal} self-interaction with a memory kernel determined by $J(\omega)$. This can be resolved with a local representation of the path integral, which can be achieved via a Hubbard-Stratonovich (HS) transformation~\cite{hubbardCalculationPartitionFunctions1959a, stratonovich1957QDistro}. This mapping allows for a return to the operator picture by replacing the nonlocal self-interaction with a local coupling to a stochastic auxiliary field. However, this transformation introduces stochasticity that must be averaged over, and it is in this averaging process that we encounter the second obstruction: the algebraic properties of the system and its coupling. While the HS transformation allows one to return to the operator picture, to obtain the exact state now requires an averaging procedure over the stochastic auxiliary field. When $[H_Q,f]\neq 0$, evaluating this average analytically presents an (apparently) \footnote{In the sequel we shall revisit and resolve this issue.} insuperable difficulty. Nevertheless, in the commuting sector $[H_Q,f]=0$, the algebraic obstruction disappears, and the averaging can be performed. Under these conditions, influence functional contains exactly the ingredients required to construct $H_{\mathrm{MF}}(\beta)$ in closed form.

Summarising, our strategy to find $H_{\mathrm{MF}}(\beta)$ is to rewrite the reduced state such that the the contributions from the bath statistics (entering only through $J(\omega)$) and the system-coupling algebra may be cleanly separated. To achieve this, it is necessary to first develop a formal treatment connecting the quenched density representation of the reduced system to a path integral representation of the influence functional.

\section{\label{sec:quenched}The Quenched Density Representation}


While many representations of reduced equilibrium exist~\cite{weissQuantumDissipativeSystems2012, breuerTheoryOpenQuantum2002, grabertQuantumBrownianMotion1988}, tracing out environmental degrees of freedom is most naturally achieved in a path integral formulation. Indeed, one of the principal advantages of this approach is that it requires no assumptions on the spectral character of the bath or the strength of its coupling to the system~\cite{weissQuantumDissipativeSystems2012}. In this framework, tracing out the bath degrees of freedom leads to the following representation of the reduced density matrix $\bar{\rho}_Q(\beta)$~\cite{feynmanQuantumMechanicsPath1965, kleinertPathIntegralsQuantum2009}:
\begin{equation}
    \langle q_f|\bar{\rho}_Q(\beta)|q_i\rangle = \frac{1}{Z_{\mathrm{tot}}} \int \mathcal{D}q \mathcal{D}\mathbf{x} \, e^{-S_E[q,\mathbf{x}]}.
\end{equation}
This is an imaginary-time path integral~\cite{feynmanQuantumMechanicsPath1965, kleinertPathIntegralsQuantum2009}, with a Euclidean action $S_E[q,\mathbf{x}]$. This contains both the bare system action, and the \emph{influence functional} $\Gamma_E[q,\mathbf{x}]$ that remains after tracing the bath degrees of freedom. While this representation makes exact treatment of the bath possible, it comes at a conceptual cost: the path
integral trades operator structure for functional structure. In particular,
questions about the \emph{operator algebra} underlying the reduced equilibrium
state are obscured once the dynamics is expressed solely in terms of paths and
kernels.

In limiting regimes the same Euclidean influence-functional machinery collapses to explicitly local reduced descriptions, e.g. in the strong-friction quantum Smoluchowski limit. This provides a useful bridge: locality can emerge either by algebraic closure (our route) or by a friction-dominated dynamical limit~\cite{AnkerholdPechukasGrabert2001}.

If one wishes to understand when a local or otherwise restricted Hamiltonian of
mean force can exist, it is therefore useful to reconnect the path-integral
description to an operator-based notion of imaginary-time propagation. The key
observation is that the reduced equilibrium operator may be regarded as an
imaginary-time evolution over an interval of length $\beta$, but with a generator
that need not be time independent. Allowing for explicit $\tau$-dependence in
the generator provides a natural bridge between the nonlocal Euclidean influence
functional and a time-ordered exponential acting on the system Hilbert space.

Motivated by this, we introduce a class of imaginary-time propagators generated by a $\tau$-dependent Hamiltonian $\bar H(\tau)$ acting on the system Hilbert
space. With this, we define the \emph{quenched} imaginary-time propagator $U(\beta)$ and the \emph{quenched density} $\rho_{\bar H}(\beta)$ as its normalised counterpart:
\begin{equation}
\begin{split}
    U(\beta)
    &:=
    \hat\tau \exp\!\left[-\int_{0}^{\beta} d\tau\,\bar H(\tau)\right],
    \\
    \rho_{\bar H}(\beta)
    &:=
    \frac{1}{Z_{\bar H}}\, U(\beta),
    \\
    Z_{\bar H}&:=\Tr\, U(\beta),
\end{split}
    \label{eq:quenched_canonical_def}
\end{equation}
where $\hat\tau$ orders increasing imaginary times (inverse temperatures) to the left~\cite{feynmanQuantumMechanicsPath1965, kleinertPathIntegralsQuantum2009}. We refer to this construction as \emph{quenched} as the $\tau$-dependence is treated as fixed
during the construction of the propagator - we may consider it as a quasistatic quench, where the Hamiltonian is modulated as temperature is lowered.

Having defined this object, we must make contact with the   path integral representation. To this end, we assume the standard kinematic structure
\begin{equation}
    \bar H(\tau)=\bar T+\bar V(\tau),
    \qquad
    \bar T=\frac{\hat p^2}{2m},
    \qquad
    \bar V(\tau)\equiv \bar V(\hat q,\tau),
    \label{eq:quenched_T_plus_V}
\end{equation}
i.e.\ the $\tau$-dependence enters through a potential term that is multiplicative in the
$|q\rangle$ basis. (This is the case relevant to the influence-functional construction
in Sec.~\ref{sec:influence_to_hmf}.)

We now discretise $[0,\beta]$ into $N$ slices of width $\Delta=\beta/N$ and define
$\tau_j=j\Delta$, $\bar H_j:=\bar H(\tau_j)$. The $\tau$-ordered exponential is then
given by the Trotter product~\cite{kleinertPathIntegralsQuantum2009,schulmanTechniquesApplicationsPath1981}
\begin{equation}
    \hat\tau \exp\!\left[-\int_{0}^{\beta} d\tau\,\bar H(\tau)\right]
    =\lim_{N\to\infty}\, e^{-\Delta \bar H_{N}}\cdots e^{-\Delta \bar H_{1}}e^{-\Delta \bar H_{0}}.
    \label{eq:quenched_timeslice}
\end{equation}
Inserting $N$ resolutions of the identity~\cite{kleinertPathIntegralsQuantum2009} in the $|q\rangle$ basis between factors, with
$q_0=q_i$ and $q_{N+1}=q_f$, we obtain the kernel
\begin{equation}
    \bar\rho(q_f,q_i;\beta)
    =
    \lim_{N\to\infty}
    \int \prod_{j=1}^{N} dq_j
    \prod_{j=0}^{N}
    \left\langle q_{j+1}\right| e^{-\Delta \bar H_j}\left|q_j\right\rangle.
    \label{eq:quenched_kernel_product}
\end{equation}

For small $\Delta$ we use a first-order Trotter splitting (sufficient in the continuum
limit)~\cite{kleinertPathIntegralsQuantum2009}
\begin{equation}
    e^{-\Delta(\bar T+\bar V_j)}
    =
    e^{-\Delta \bar T}\,e^{-\Delta \bar V_j}
    +\mathcal O(\Delta^2),
    \qquad
    \bar V_j:=\bar V(\hat q,\tau_j),
    \label{eq:trotter_split}
\end{equation}
so that
\begin{equation}
    \langle q_{j+1}|e^{-\Delta \bar H_j}|q_j\rangle
    =
    \langle q_{j+1}|e^{-\Delta \bar T}|q_j\rangle\;
    e^{-\Delta \bar V(q_j,\tau_j)}
    +\mathcal O(\Delta^2).
    \label{eq:short_time_kernel_split}
\end{equation}

The kinetic matrix element is evaluated by inserting a momentum resolution of
the identity~\cite{feynmanQuantumMechanicsPath1965,schulmanTechniquesApplicationsPath1981}:
\begin{align}
    \langle q_{j+1}|e^{-\Delta \bar T}|q_j\rangle
    &=
    \int \frac{dp}{2\pi}\,
    \langle q_{j+1}|p\rangle\langle p|q_j\rangle\,e^{-\Delta p^2/(2m)}
    \nonumber\\
    &=
    \sqrt{\frac{m}{2\pi\Delta}}\,
    \exp\!\left[-\frac{m}{2\Delta}(q_{j+1}-q_j)^2\right].
    \label{eq:kinetic_short_time}
\end{align}
\begin{widetext}
Substituting Eq.~\eqref{eq:kinetic_short_time} and Eq.~\eqref{eq:short_time_kernel_split} into
Eq.~\eqref{eq:quenched_kernel_product} yields
\begin{equation}
    \bar\rho(q_f,q_i;\beta)
    =
    \lim_{\Delta\to 0}
    \int \prod_{j=1}^{N} dq_j
    \prod_{j=0}^{N}
    \left(\sqrt{\frac{m}{2\pi\Delta}}\right)
    \exp\!\left[
        -\sum_{j=0}^{N}
        \Delta\left(
            \frac{m}{2}\left(\frac{q_{j+1}-q_j}{\Delta}\right)^2
            +\bar V(q_j,\tau_j)
        \right)
    \right].
    \label{eq:quenched_discrete_PI}
\end{equation}

In the limit $N\to\infty$ (i.e.\ $\Delta\to 0$), Eq.~\eqref{eq:quenched_discrete_PI} becomes the
Euclidean path integral~\cite{feynmanQuantumMechanicsPath1965,kleinertPathIntegralsQuantum2009,groscheHandbookFeynmanPath1998}, such that the reduced density (normnalised by $Z_{\bar H}$) is given by
\begin{equation}
    \rho_{\bar H}(q_f,q_i;\beta)
    =
    \frac{1}{Z_{\bar H}}
    \int_{q(0)=q_i}^{q(\beta)=q_f}\!\!\mathcal D q(\tau)\;
    \exp\!\left[
        -\int_0^\beta d\tau\,
        \left(
            \frac{m}{2}\dot q(\tau)^2+\bar V(q(\tau),\tau)
        \right)
    \right].
    \label{eq:quenched_continuum_PI}
\end{equation}
\end{widetext}
The Euclidean action $S[\bar q]=\int_0^\beta d\tau\,\big[\frac{m}{2}\dot q^2+\bar V(q,\tau)\big]$
appearing in the exponent defines the system action referred to in subsequent
sections. We note that for finite-dimensional systems, while the specific form of the path integral measure changes (e.g., to a sum over discrete trajectories or coherent states), the quenched density representation retains precisely the same form~\cite{weissQuantumDissipativeSystems2012, altlandCondensedMatterField2010}. The separation into a bare system action and a bath-induced stochastic driving is a consequence of the bath properties alone, and does not depend on the dimensionality of the system Hilbert space~\cite{leggettDynamicsDissipativeTwostate1987}.
\section{\label{sec:influence_to_hmf}From Influence Functionals to the Hamiltonian of Mean Force}
Having established the equivalence between the quenched density and the standard Euclidean path integral in the previous section, we now outline the specific form of the reduced system when coupled to a harmonic bath. Following the influence-functional formalism \cite{feynmanTheoryGeneralQuantum1963a,caldeiraQuantumTunnellingDissipative1983a,grabertQuantumBrownianMotion1988} and specifically its stochastic representation for Gaussian baths \cite{mccaulPartitionfreeApproachOpen2017c, mccaulDrivingSpinbosonModels2018a, mccaulHowWinFriends2021b}. The resulting reduced density matrix admits the imaginary-time path integral representation:
\begin{widetext}
\begin{equation}
    \bar\rho_Q(q_f,q_i;\beta) = \int_{q(0)=q_i}^{q(\beta)=q_f}\!\!\mathcal D q\; \exp\Bigg[ -S_Q[q] + \frac{1}{2}\int_0^\beta d\tau\int_0^\beta d\tau' f\big(q(\tau)\big) K(\tau-\tau') f\big(q(\tau')\big) \Bigg].
    \label{eq:reduced_effective_PI}
\end{equation}
\end{widetext}
Here $Z_X(\beta)=\Tr_X e^{-\beta H_X}$ is the free bath partition function, $S_Q$ is the Euclidean system action defined and $K(\tau)$ is the imaginary-time influence kernel. In terms of the bath spectral density $J(\omega)$, the kernel is given by~\cite{weissQuantumDissipativeSystems2012, feynmanTheoryGeneralQuantum1963a}
\begin{equation}
    K(\tau) = \frac{1}{\pi}\int_0^\infty d\omega\, J(\omega) \frac{\cosh\!\left[\omega\left(\frac{\beta}{2}-|\tau|\right)\right]} {\sinh\!\left(\frac{\beta\omega}{2}\right)}.
    \label{eq:kernel_spectral}
\end{equation}
Related exact Gaussian-bath embeddings that trade nonlocal memory for time-local dynamics in a minimally extended state space are developed in Ref.~\cite{XuAnkerholdEPJST2025}; here we instead emphasise the Euclidean equilibrium generator and its exact closure conditions.

While the representation in Eq.\eqref{eq:reduced_effective_PI} is exact, it is nonlocal (bilocal) in imaginary time. This nonlocality can be resolved by a Hubbard - Stratonovich (HS) transformation \cite{hubbardCalculationPartitionFunctions1959a, stratonovich1957QDistro}, which decouples the bilocal interaction by introducing an auxiliary Gaussian field $\xi(\tau)$. This step provides the bridge to an operator-level representation of $\bar{\rho}_Q$ as a Gaussian average over the local quenched canonical densities discussed in Sec.~\ref{sec:quenched}:
\begin{equation}
    \bar{\rho}_Q(\beta) = \mathbb{E}_\xi\bigg[\hat{\mathcal{T}}\exp\!\left(-\int_0^\beta d\tau\,\bar{H}(\tau;\xi)\right) \bigg],
    \label{eq:bar_rho_as_quenched_average}
\end{equation}
where $\bar{H}(\tau;\xi) = H_Q - \xi(\tau)f$ is the time-dependent quenched Hamiltonian.
The statistical properties of the auxiliary field $\xi(\tau)$ are set by the influence kernel \cite{mccaulPartitionfreeApproachOpen2017c,mccaulDrivingSpinbosonModels2018a,mccaulHowWinFriends2021b}
\begin{equation}
    \mathbb{E}_\xi[\xi(\tau)] = 0, \quad \mathbb{E}_\xi[\xi(\tau)\xi(\tau')] = K(\tau-\tau').
    \label{eq:noise_statistics}
\end{equation}
This Hubbard--Stratonovich unravelling is the same as that underlying partition-free stochastic Liouville formulations of Gaussian environments and equilibrated initial conditions~\cite{mccaulPartitionfreeApproachOpen2017c}. Recent work also develops a time-local minimally extended-state-space embedding (QD-MESS) via influence-functional decomposition and Hubbardâ€“Stratonovich unravelling, recovering Caldeiraâ€“Leggett and Smoluchowski-type limits as special cases~\cite{XuAnkerholdEPJST2025}.

Having performed this transformation, the operator representation maps directly to the quenched density framework. The challenge now lies in performing the average over $\xi$. To do so, we must isolate the stochastic term. Moving to the interaction picture with respect to $H_Q$, the propagator factorises as
\begin{equation}
    \hat{\mathcal{T}}e^{-\int_0^\beta d\tau\,[H_Q - \xi(\tau)f]} = e^{-\beta H_Q} \hat{\mathcal{T}}e^{\int_0^\beta d\tau\, \xi(\tau) f(\tau)},
    \label{eq:interaction_picture_split}
\end{equation}
where $f(\tau) = e^{\tau H_Q} f e^{-\tau H_Q}$. While we have successfully isolated the stochastic field $\xi(\tau)$, the coupling operator $f(\tau)$ has acquired a nontrivial imaginary-time dependence. This greatly complicates the averaging out of the stochastic term, and we defer its explicit evaluation to future work. 

Here we focus instead on the commuting (QND) sector $[H_Q,f]=0$, where the algebraic obstruction disappears: $f(\tau)=f$ is $\tau$-independent, the $\hat{\mathcal{T}}$ ordering becomes redundant and the expression factorises:
\begin{align}
    \hat{\mathcal{T}}\exp\!\left[-\int_0^\beta d\tau\,\bar{H}(\tau;\xi)\right]
    &=
    e^{-\beta H_Q}\, \exp\!\left[f\,X\right],
    \label{eq:tilde_rho_commuting_factorised}
\end{align}
where $X:=\int_0^\beta d\tau\,\xi(\tau)$ is a Gaussian variable with mean zero and variance
\begin{equation}
    C(\beta) = \mathbb E_\xi[X^2] = \int_0^\beta d\tau\int_0^\beta d\tau'\,K(\tau-\tau').
    \label{eq:Cbeta_def}
\end{equation}
Substituting the kernel Eq.~\eqref{eq:kernel_spectral} and performing the double
imaginary-time integral gives
\begin{align}
    C(\beta)
    &=\frac{2\beta}{\pi}\int_0^\infty d\omega\,\frac{J(\omega)}{\omega}
    \equiv 2\beta\,\lambda,
    \nonumber\\
    \lambda&:=\frac{1}{\pi}\int_0^\infty d\omega\,\frac{J(\omega)}{\omega},
    \label{eq:Cbeta_explicit}
\end{align}
where $\lambda$ is the reorganisation energy~\cite{weissQuantumDissipativeSystems2012, leggettDynamicsDissipativeTwostate1987}.
Crucially, $C(\beta)$ is \emph{linear} in $\beta$, so the ratio $C(\beta)/(2\beta)=\lambda$ is temperature independent.

With this the Gaussian average $\mathbb E_\xi[e^{f X}] = \exp[\frac{1}{2}C(\beta)f^2]$ can be performed exactly, and using Eq.~\eqref{eq:bar_rho_as_quenched_average}, we obtain:
\begin{equation}
    \bar\rho_Q(\beta) = e^{-\beta H_Q}\,\exp\!\left[\beta\lambda\,f^2\right].
    \label{eq:bar_rho_commuting_closed}
\end{equation}

This is (up to normalisation) the canonical exponential form for a thermal state, and the Hamiltonian of mean force is identically
\begin{equation}
    H_{\mathrm{MF}}(\beta) = H_Q -\lambda\,f^2 
    \label{eq:HMF_commuting_final}
\end{equation}
The only nontrivial operator content is the correction $-\lambda f^2$, which is temperature independent.

An immediate consequence that can be inferred from this result is that it is identical to the result in the classical mean-force construction
\cite{campisiTalknerHanggi2009Solvable,talknerColloquiumStatisticalMechanics2020,cerisolaQuantumClassicalCorrespondence2024}. This equilvance can be understand mostly easily through the link to Koopman von-Neumann dynamics. In real time, classical evolution is generated by the Liouville/Koopman operator $\mathcal L_H=\{\,\cdot\,,H\}$, with propagator $e^{t\mathcal L_H}$. To obtain Gibbs states from imaginary time however, the relevant generator becomes the Hamiltonian. This is true in both the quantum and classical scenarios, but in the latter case all observable operators (and hence the Hamiltonian) commute identically. The classical equilibrium construction will therefore always lie within the commuting sector of the fully quantum problem. For a fuller discussion of the dynamical distinctions between quantum and classical systems in both real and imaginary time, see ref.~\cite{McCaulZhdanovBondar2023WaveOperator}.

Another slight (but interesting) consequence of this result concerns integrable systems where the coupling operator $f$ is a conserved quantity ($[H_Q, f]=0$). In this case, the Hamiltonian of mean force $H_{\mathrm{MF}}$ shares the same eigenbasis as the bare system Hamiltonian. The strong coupling to the bath therefore does not mix the system energy levels, but rather induces a level-dependent energy shift $-\lambda f_n^2$, where $f_n$ are the eigenvalues of the coupling operator. The integrability of the open system is thus preserved in the sense that the equilibrium state remains diagonal in the energy basis of the isolated system, but depending on the specific structure of $f$, the bath can induce a non-trivial operator-level effect on the reduced equilibrium state. 

A further important consequence is that $\lambda$ coincides with the static potential renormalisation $\frac{1}{2}\sum_k c_k^2/(m_k\omega_k^2)$ of the Caldeira-Leggett model. If the standard counterterm $+\lambda f^2$ has been absorbed into $H_Q$ (as stated in Sec.~\ref{sec:model}), the bath-induced shift $-\lambda f^2$ cancels it exactly. In this case, $H_{\mathrm{MF}}$ reduces to the bare system Hamiltonian $H_Q^{\mathrm{bare}}$, and the normalised mean-force Gibbs state coincides with the bare-system Gibbs state. In other words, in the commuting sector with translational invariance enforced, the bath has no operator-level effect on the reduced equilibrium
state. Conversely, if no counterterm is included, $-\lambda f^2$ represents a genuine
bath-induced static renormalisation of the system potential. Amusingly, this result closely aligns to the master equation folk wisdom where superior results can be obtained when one neglects \emph{both} counterterms and Lamb-shifts~\cite{correaPotentialRenormalisationLamb2025}.

Having established the quenched density representation exactly captures the reorganisation-energy result~\cite{campisiTalknerHanggi2009Solvable}, we now apply it to stochastic numerical evaluation of the equilibrium state.

\section{Numerical Evaluation}
\label{sec:numerical}
We now verify the predicted $H_{\mathrm{MF}}$ can be obtained numerically through stochastic averaging. To do so, we consider a toy five-level system $H_S = \sum_{n=0}^4 E_n |n\rangle\langle n|$ coupled to the bath via the projector $f = |4\rangle\langle 4|$. This coupling is manifestly commuting ($[H_S, f] = 0$), ensuring that the exact HMF should be
\begin{equation}
    H_{\mathrm{MF}} = H_S - \lambda |4\rangle\langle 4| + \mathrm{const}.
\end{equation}
The precise level spacings are selected randomly and kept fixed across all $(\beta,g)$ simulations. For the bath, we set use $J_g(\omega)=2 g^2\omega e^{-\omega/\omega_c}$, incorporating system-bath coupling strength directly into the parameter $g$. We choose $\omega_c=1$ for all simulations. As a point of comparison, we further implement an explicit bath of $K$ harmonic modes $a_k$, coupling to the system through $\sum_k a_k +a_k^\dagger$. The bath spectrum is then mimicked by discretising the interval $\omega\in[0, \omega_c]$ into mode bins with centres $\omega_k$ and widths $\Delta\omega_k$, setting $c_k^2=\frac{2}{\pi}J_g(\omega_k)\,\omega_k\,\Delta\omega_k$. This ensures that the discretised reorganisation scale $\lambda_{\mathrm{disc}}=\sum_k\frac{c_k^2}{2\omega_k^2}$ tracks the continuum expression $\lambda_{\mathrm{cont}}=\frac{1}{\pi}\int_0^\infty\frac{J_g(\omega)}{\omega}\,d\omega$~\cite{weissQuantumDissipativeSystems2012}. This provides an exact reference for the reduced state $\rho_Q^{\mathrm{ED}}=\Tr_B[e^{-\beta H_{\mathrm{tot}}}]/\Tr[e^{-\beta H_{\mathrm{tot}}}]$. 

\begin{figure}[htbp]
\includegraphics[width=\columnwidth]{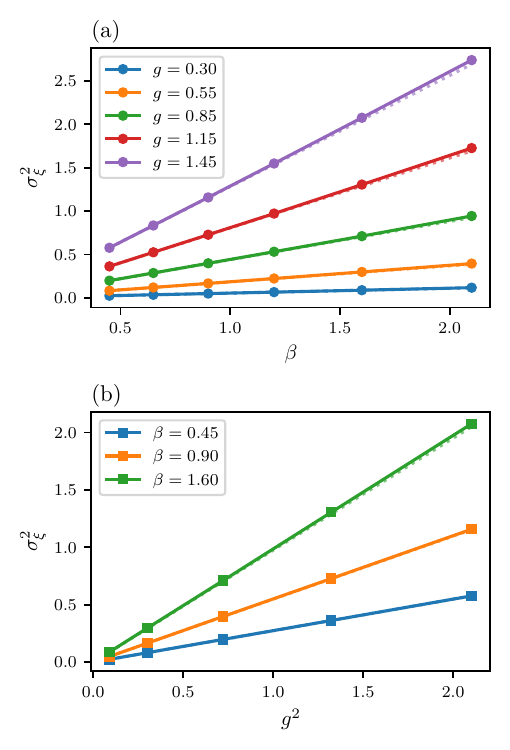}
\caption{\label{fig:variance_scaling} Variance scaling. The sample variance $\sigma^2_\xi$ of the stochastic field grows linearly with inverse temperature $\beta$ (a) and linearly with the reorganisation energy $\lambda \propto g^2$ (b). Symbols are simulation results; dotted lines are the theoretical prediction $2\beta\lambda_{\mathrm{disc}}$.}
\end{figure}

Finally, we implement the stochastic formulation of the quenched density given by Eq.~\ref{eq:bar_rho_as_quenched_average}). The Gaussian noise $\xi(\tau)$ is generated by convolution of white noise with the covariance matrix of the discretised kernel $K(\tau)$, a standard technique for sampling coloured noise (see e.g. \cite{mccaulDrivingSpinbosonModels2018a}). In the present model, the variance of the integrated field $X=\int_0^\beta d\tau\,\xi(\tau)$ is predicted to be $2\beta\lambda_{\mathrm{disc}}$. Fig.~\ref{fig:variance_scaling} confirms this scaling, plotting the sample variance extracted from the stochastic simulations against $\beta$ (panel a) and $g^2$ (panel b).

\begin{figure}[htbp]
\includegraphics[width=\columnwidth]{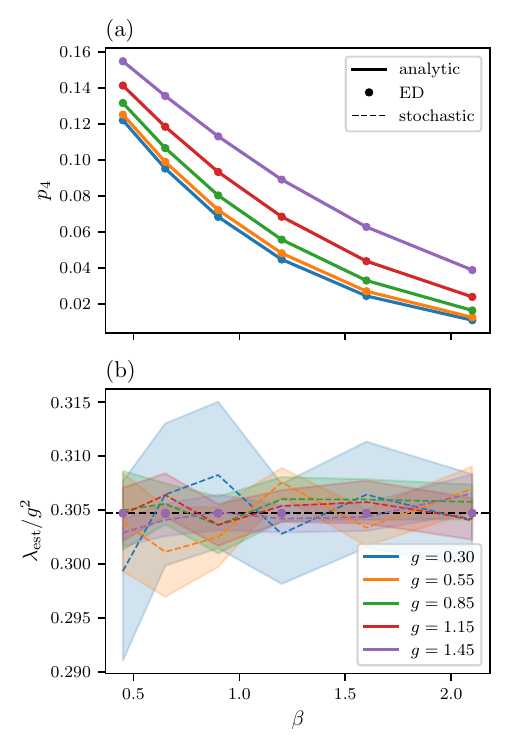}
\caption{\label{fig:temp_dep} Temperature dependence and scaling. (a) Population of the coupled state ($|4\rangle$) vs $\beta$ for various couplings $g$. Circles (ED) match solid lines (analytic). (b) The extracted reorganisation energy $\lambda_{\mathrm{est}}$ collapsed by $g^2$ vs $\beta$. All curves hover near the theoretical thermodynamic limit (dashed lines), verifying that the bath's effect reduces to a single reorganisation energy $\lambda$.}
\end{figure}

With this setup the core predictions of the HMF framework can be verified. 
Fig.~\ref{fig:temp_dep} confirms the thermodynamic consistency of the quenched density framework, showing that the effective Hamiltonian correctly captures the temperature dependence of the open system, with the extracted reorganisation energy matching the theoretical value. 

\begin{figure}[htbp]
\includegraphics[width=\columnwidth]{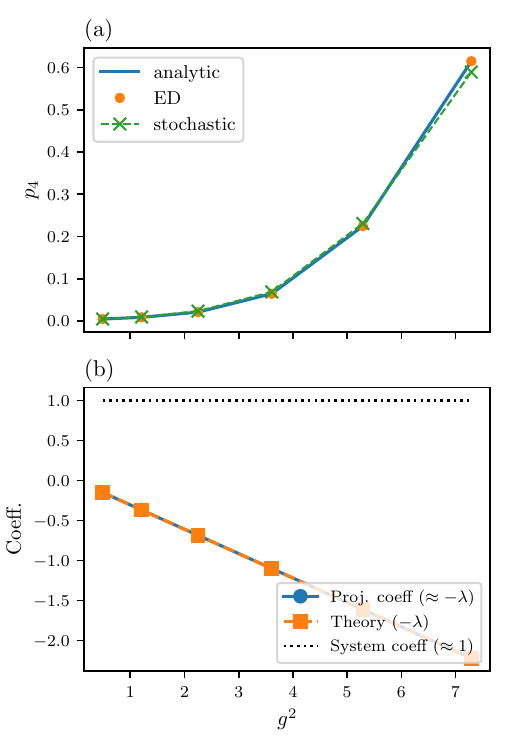}
 \caption{\label{fig:strong_coupling} Strong-coupling benchmark. (a) Population $p_4$ vs $g^2$ at fixed $\beta$, showing excellent agreement over the full range. (b) Coefficients in the system basis vs $g^2$. The projector component $P_4=|4\rangle\langle 4|$ scales as $-\lambda$, while off-diagonal coherences are exponentially suppressed. This strong-coupling projector dominance is consistent with ultrastrong-coupling analyses of the mean-force Gibbs state~\cite{cresserWeakUltrastrongCoupling2021a}. The bare system Hamiltonian $H_S$ remains unrenormalised (dotted), confirming $H_{\mathrm{MF}} = H_S - \lambda f^2$.}
\end{figure}

Fig.~\ref{fig:strong_coupling} demonstrates robustness in the strong-coupling regime, where the interaction strength rivals the system energy scales. To verify the operator structure of the HMF, we fit the numerically extracted $H_{\mathrm{MF}}$ to the ansatz $c_0 \mathbb{I} + a_{H_S} H_S + a_P P$. As shown in panel (b), the coefficient $a_P$ tracks the theoretical prediction $-\lambda$ perfectly, while $a_{H_S} \approx 1$ confirms that the bare system Hamiltonian is not renormalised. 
 
\begin{figure}[htbp]
\includegraphics[width=\columnwidth]{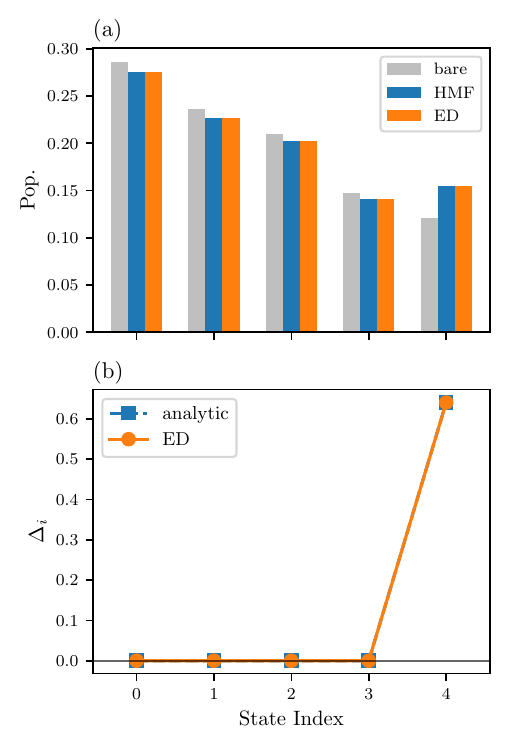}
\caption{\label{fig:microscopic} Microscopic validation. (a) Detailed population snapshot at fixed $\beta, g$: "Analytic HMF" (blue) corrects the "Bare Gibbs" (grey) to match ED (orange). (b) Effective energy shifts $\Delta_i$ for each level; only the coupled level ($|4\rangle$) shifts, by exactly $-\lambda$. Non-coupled levels remain unperturbed.}
\end{figure}

Finally, Fig.~\ref{fig:microscopic} provides the microscopic resolution: the HMF modifies only the coupled subspace, leaving the orthogonal manifold untouched, as predicted by the operator structure of Eq.~\ref{eq:HMF_commuting_final}.

\section{Discussion and outlook}
\label{sec:discussion}

We have presented the quenched density framework as a rigorous bridge between path-integral influence functionals and operator-level thermodynamics. By mapping the bath trace to a stochastic average over quenched propagators, we isolate the algebraic structure of the system's response from the statistical properties of the environment.

In the case of harmonic baths where $H_Q$ and $f$ commute, the algebraic and statistical contributions to the HMF can be cleanly partitioned: the bath statistics collapse to a scalar reorganisation energy $\lambda$, and the system algebra reduces to a static potential shift $-\lambda f^2$. Our numerical benchmarks (Sec.~\ref{sec:numerical}) confirm that this framework reproduces the exact finite-bath physics, regardless of coupling strength.

The exact result derived here is not novel, and in isolation can hardly justify the formal machinery used to derive it. In this regard, our treatment is incomplete. The power of the quenched density lies in its modularity, which naturally accommodates both more general environments, and system bath-couplings. In the latter case, the challenge shifts from the statistical generation of noise to the Lie-algebraic properties of the time-ordered effective propagator. The present framework handles this by clearly delineating the statistical complexity of the bath from the dynamical complexity of the system response, meaning it is possible - though nontrivial - to incorporate both non-commuting couplings and anharmonic baths. In forthcoming work, we shall address these generalisations. Even in this simplified case however, we see the shape of the answer to the question with which we began. What governs equilibrium? Algebra.

%

\end{document}